\documentstyle[12pt,aasms4]{article}

\lefthead{Aller et al.}
\righthead{Radio Properties of BL~Lacs}

\begin{document}

\title{Cm-Wavelength Total Flux and Linear Polarization Properties of
 Radio-Loud BL~Lacertae Objects}
\author{Margo F. Aller, Hugh D. Aller, Philip A. Hughes, and George E. Latimer}
\affil{Department of Astronomy, University of Michigan, Dennison Building, Ann
Arbor, MI 48109-1090}
\authoremail{margo@astro.lsa.umich.edu, hugh@astro.lsa.umich.edu, hughes@astro.lsa.umich.edu, gel@astro.lsa.umich.edu}

\begin{abstract}

We present results from a long-term program to quantify the range of
behavior of the cm-wavelength total flux and linear polarization variability properties of a sample of 41 radio-loud BL~Lac objects using weekly
to tri-monthly observations with the University of Michigan 26-meter
telescope operating at 14.5, 8.0, and 4.8~GHz; these observations are used to
identify class-dependent differences between these BL~Lacs and QSOs 
in the Pearson-Readhead sample. As a group, the BL~Lacs are found to be more
highly variable in total flux density than the QSOs. These changes are often
nearly-simultaneous and of comparable amplitude at 14.5 and 4.8~GHz, which
contrasts with the behavior in the QSOs and supports the existence of class-dependent differences in opacity within the parsec-scale jet flows. Structure function analyses of the flux observations quantify that a characteristic timescale is identifiable in only 1/3 of the BL~Lacs and that
in the majority of the program sources the activity is uncorrelated within the
timescales probed. The time-averaged fractional linear polarizations are only
on the order of a few percent and are consistent with the presence of
tangled magnetic fields within the emitting regions. In many sources a 
preferred long-term orientation of the electric vector position angle is present. When compared with the VLBI structural axis, no  preferred position angle difference is identified; the derived distribution resembles that known
for core components from VLBP measurements. The polarized flux typically
exhibits variability with timescales of months to a few years and exhibits the signature of a propagating shock during several resolved outbursts. The flux
and polarization variability indicate that the source emission is predominately 
due to evolving source components and supports the occurrence of more frequent shock formation in BL~Lac parsec-scale flows than in QSO jets where 
the magnetic field topology even during outbursts is similar to that 
of the underlying quiescent flow. The differences in variability behavior
and polarization between BL~Lacs and QSOs which we find can be explained by
differences in stability between the jet flows found by recent
studies of relativistic hydrodynamic flows.

\end{abstract}


\section{Introduction}

 Several lines of evidence based on radio data have supported the view that the
physical conditions in the jets of BL~Lacertae type objects, an
optically-defined class, are intrinsically different from those found in QSO
and galaxy jets, and that these differences cannot be accounted for
solely by differences in the viewing angle to the flow axis. The extensive VLBP studies of Gabuzda and co-workers ({\it e.g.} Gabuzda \& Cawthorne 1992), based
primarily on observations at 5 GHz, for example, have
shown that the magnetic fields in BL~Lac jet components are characteristically
aligned perpendicular to the direction defined by source structure at
cm-wavelengths rather than along the structural axis, an orientation found
typically for QSOs. The broadband study by Sambruna et al. 
(1996) which documented the distinct shape and position of the
turnover in the radio to X-ray spectral energy distribution for several types
of blazars, has also shown
that class differences cannot easily be explained by viewing angle alone; comparison of these data with jet models demonstrated that differences in intrinsic
source parameters are required to explain the differences in shape of 
the spectral energy distributions. Recent VLBI studies, however, including observations at shorter wavelengths, have identified exceptions to a simple
VLBP morphological class separation ({\it e.g.} Gabuzda 1995; Lister, Marscher,
\& Gear 1998).

In 1979, as part of an effort to study the radio properties of class members
and to identify class-dependent differences in the parsec-scale flows
compared to other AGNs, we initiated a program to systematically
observe a sample of BL~Lac objects with the University of Michigan 26-meter paraboloid.  A goal was to obtain data with frequent time resolution to
quantify the behavior of the centimeter-wavelength total flux density and linear polarization originating in the inner jets of these sources. The sample is complete to 0.4 Jy, with sampling sufficient to permit detection of variations with timescales of weeks or longer.  The results obtained from this longterm program are discussed here. In section 2 we present the sample 
and describe selection criteria. In sections 3 and 4 we summarize the variability, spectral, and polarization properties of the sample members, compare the statistical properties of BL~Lacs and QSOs, and discuss the
relation of our data to results from VLBI/P studies. In section 5 
we summarize the evidence from our variability program supporting intrinsic class differences and discuss these results in the context of recent
hydrodynamic studies of jet flows. Section 6 presents the conclusions 
resulting from our study and suggests future work.

\section{Definition of the Sample}

Our program to systematically study the radio properties of a well-defined
group of BL~Lac objects commenced in August 1979; a few of the objects had 
been observed prior to this time as part of the Michigan (hereafter UMRAO)
variability program. Initially, the objects in a list of all known BL~Lacs supplied by J. Ledden (private communication), to appear shortly thereafter
in slightly modified form in Hewitt \& Burbidge (1980), were observed at 8.0
GHz in a series of runs in August and December 1979 to identify class members sufficiently strong for inclusion; this list included objects which are highly compact, vary in radio, optical or X-ray wavelengths, are often highly polarized, and exhibit either no spectrum lines or only very weak features in the optical spectral region. These conditions for BL Lac membership do not
include a specific requirement on line width but add a requirement of variability which is not included in some compilations of BL Lac members. The
additional criteria which we adopted for program membership are:
1) S$>0.4$ Jy at 8 GHz in August-December 1979 and 2) $-36 \arcdeg \leq \delta \leq\ +81 \arcdeg$.
The limit on total flux density was selected to permit sufficient signal-to-noise in the measurement of the linear polarization (typically only
a few percent of the total flux density) for the analysis, and the restrictions on the declination range were set by the mechanical limits of the Michigan paraboloid. In early 1980 the source 1413+135, not on the Ledden list, was 
added to the program, while in mid 1980 through early 1981 a few newly-identified class members, primarily high declination objects  from the Bonn-NRAO 5~GHz survey (Biermann  et al. 1981), were also added, and a few of the original objects were removed because of reclassification. In 1984 the
Pearson-Readhead (1981, 1988; hereafter PR) survey member 0954+658 was also added; the sources 0814+425 and 1823+568 classified as BL~Lacs in PR were not added because they did not meet the adopted class definition (Burbidge \& Hewitt 1992). Observations of the source 3C~66A, which meets our selection criteria,
are not included in this study because the source emission is confused within
the beam of our paraboloid with 3C~66B, and, also, measurements of the low
declination source 2155$-$304, just at our flux limit and in a region of the 
sky with reduced pointing accuracy, were excluded because of poor signal-to-noise. The final list of 41 program objects is given in Table~1.  All sample members meet the Hewitt and Burbidge definition for class membership described above (Hewitt \& Burbidge 1980; Burbidge \& Hewitt 1987). Because of consistency in classification criteria, they represent a homogeneous group for
investigating the range of behavior in BL~Lacs. 

Selected information on the broadband emission properties of the sources is presented in Table 1. Some objects at times exhibit broad emission-line spectra
or, in some cases, absorption lines typical of QSO-type spectra (Burbidge \&
Hewitt 1992; Vermeulen et al. 1995), and these are distinguished by an `x' in
column 3. We also include 0306+102 in this group; 
based on its spectral features, V\'eron (1994) has suggested that it is a
QSO. Column 4 lists values of z taken from the literature; the sample contains
4 very low z (z $\leq 0.050$) objects and includes objects in the range $0.03 \leq $ z  $ \leq 1.72$.  In column 5 we include information on membership in widely-referenced data compilations: the Stickel et al. (1991) 1 Jy radio 
survey (based on the 5 GHz flux in K\"uhr et al. 1981), the Einstein slew
survey (Perlman et al. 1996); and objects detected by EGRET (Mattox et al. 1997). Inclusion in these radio, X-ray and gamma-ray studies is indicated by 
R, X, and G respectively. Column 6 gives the average flux density at 14.5~GHz.
These values are typically only on the order of 1 Jy, although fluxes can
increase by an order of magnitude during active periods. 

Program objects were observed {\it at least} tri-monthly at our three observing
frequencies (4.8, 8.0, and 14.5~GHz) following the data acquisition,
reduction, and calibration procedures described in Aller, Aller, Latimer, \&
Hodge (1985). Objects identified as currently variable were incorporated into
the UMRAO core program of active objects for weekly or more-frequent observation; thus, during active phases the objects were observed intensively, while the sampling is relatively sparse during quiescent phases. Data through 1 September 1996, including 9,827 observations at 14.5~GHz and 6,186 observations at 4.8~GHz, are discussed here. 

\section{Total Flux Variability}

To parameterize the characteristic behavior of the flux variability, we have
computed both a variability index measuring the peak-to-trough amplitude change and a characteristic cm-wavelength spectral index, and identified a
characteristic timescale of the variability from a first-order structure function analysis of our data.  As a simple measure of the amplitude change in total flux, we define a variability index which measures the peak-to-trough variation during the time window of the observations: 
\begin{equation}
V = \frac{S_{max}-S_{min}}{S_{max}+S_{min}}.
\end{equation}
Daily averages of the data were used to identify peak and trough fluxes
for each source. Spectral indices were computed from paired values of
monthly averages of the data at 4.8 and 14.5~GHz; monthly averages were used 
to ensure a time match between the two frequencies for the most poorly 
observed sources when following the spectral development. We adopt the sign convention that $S\propto\nu^{+\alpha}$. In computing these parameters
characterizing the source emission, observations with $\sigma_{S}>$ max(0.1Jy,
0.03S) were excluded to remove observations with unusually poor 
signal-to-noise.  The identification of the characteristic
timescale from the turnover in a plot of log(SF) versus log(lag) follows 
the procedure described in Hughes, Aller, \& Aller (1992). Although some 
sources had been observed prior to the beginning of this study, only data
during the common time period 1980.0 to 1996.7 are included in the computation of the variability and spectral indices and the timescales.

 Columns 7-12 of Table 1 summarize the results. Columns 8 and 10 give
the number of observations at 14.5 and 4.8~GHz respectively, meeting the inclusion criteria described above and used to determine the values of V given
in columns 7 and 9. We note that the amplitude change measured by V
can be sensitive to the choice of the time window; in several sources a large value reflects the occurrence of a single unusually-large-amplitude flux
change relative to those more commonly seen. Such large events are typically separated by 10-15 years. Column~11 contains the average of the spectral
indices computed from monthly averages of the data; this parameter is an
indicator of the global flatness of the spectrum. Column~12 contains the
maximum correlation timescale in years found from  first-order structure function analyses of the data.  Only 1/3 of the BL~Lacs show a well-defined plateau in their structure functions; of these OJ~287 exhibits a double
plateau. For the remainder, a characteristic timescale could only be poorly
to very poorly determined (marked by `*' and `?' respectively). In some cases $\sigma_{SF}$ was large because of sparse (trimonthly) sampling.  In 3 sources -- the intraday variable 0716+714, and the 2 relatively weak sources 2032+107 and 2335+031 -- no value of the timescale could be measured. The structure
function analysis indicates that the cm-wavelength variations in the majority
of the BL~Lacs in our current unbiased sample are uncorrelated for the range 
of timescales probed by the
data. This result contrasts with that presented in Hughes, Aller, \& Aller (1992) which was based on a study of a nonhomogeneous sample composed of the best-observed sources in our monitoring program; there a well-defined 
plateau in the structure function was found for {\it nearly all} the sources
studied. However, these sources were selectively chosen for
frequent observation because they were known to exhibit large-amplitude,
relatively well-resolved outbursts; they are not typical of AGNs, but rather
the most extreme examples. Nineteen sources are contained in both
studies; of these, two sources which previously exhibited variability
characteristic of white noise (Mrk~501 and 3C~371) now have defined timescales,
identified because of the increased length of the time window. In a few
objects differences were found in the timescale determinations which are
consistent with the time-dependent changes in variability apparent in the light
curves; such changes in behavior provide evidence for a non-stationary process in these sources.

While as a group the BL~Lacs exhibit large amplitude changes in total flux
over the full time period studied, as shown by the values of V, the detailed
structure of the variability can be very different from source to source. For
example, objects such as 0235+164, OJ~287, 1418+546, and BL~Lac exhibit rapid
well-resolved outbursts characterized by sharp rises and declines (see {\it
e.g.} Aller, Aller, \& Hughes 1996b); others show long-term (up to decade-long)
rises or declines, on which shorter-term variations may also be superimposed.
This range in behavior is illustrated in Figures 1 and 2 which show the data for the very rapid variable 0235+164 and for the slow variable ON~231; the characteristic time scales for the flux variations, in the source frame, found from the structure function analysis are 0.5 years and 5.5 years respectively.
In 0235+164 the activity is nearly continuous with events that are
well-resolved in time. No strong evidence has been found for periodicity 
in this source or others in our program, with the exception of OJ~287 (Hughes, Aller, \& Aller 1998). The behavior in ON~231 is characterized by a long-term decline on which some structure is superimposed. While 0235+164 has been found by most VLBI investigators to be unresolved (Gabuzda \& Cawthorne 1996), VLBI measurements of ON 231 reveal a complex structure with many components 
(Gabuzda et al. 1994a) whose contributions are blended in our integrated total flux measurements. Note that in these two objects, as in the majority of our 
program BL~Lacs, the spectra remain flat even during outburst development
and that {\it no} quiescent periods are observed.

A goal of this study is to compare the variability behavior of BL~Lacs with
that of QSOs. Because the Michigan variability program is directed towards
observing the most active sources, results based on the complete UMRAO database yield a biased representation of the general behavior of
extragalactic objects. A useful comparison, however, can be made using
the subsample of QSOs in the flux-limited Pearson-Readhead (1981, 1988) survey which we have been observing since 1984 (Aller, Aller, \& Hughes 1992). 
The result of this comparison for the data at 14.5~GHz is illustrated in Figures 3 and 4 which show histograms of the distribution of the variability index for the two samples based on data during the common observing time period 
for the two samples -- 1984.0 to 1996.7. Corresponding histograms for the data
at 4.8~GHz for the two source groups are shown in Figures 5 and 6. In Figures 
3 and 5 the BL~Lacs have been coded to indicate those objects which sometimes 
show QSO-like properties; in Figures 4 and 6, following the suggestion by Valtaoja and coworkers ({\it e.g.} Valtaoja 1994) that higher amplitude variability occurs in HPQs than in LPQs, the PR~QSOs have also been
separated into two groups: LPQ ($<3\%$) and HPQ ($>3\%$) based on their
maximum-observed optical polarization, given in  Impey, Lawrence, \& Tapia (1991) supplemented by Wills et al. (1992). For three weak QSOs (1624+416, 1634+628 and 2351+456), indicated as NCl, no optical polarization data were
obtained in these studies, and they are unclassified. It should be noted that
for many of these QSOs the optical polarization data have been only very
sparsely sampled; thus, the range of the variability in optical percentage polarization has not been well-determined. Table~2 summarizes the range and distribution of the variability indices for the BL~Lacs and PR~QSOs. The distributions indicate that the BL~Lacs as a group are {\it highly} variable
relative to the QSOs, even when compared to the more variable HPQs. A KS test 
of the similarity of paired distributions indicates a probability of $1.87
\times 10^{-4}$ that the distributions of V at 14.5~GHz for the two samples
are drawn from the same population and a probability of only $1.0 \times
10^{-6}$ that the distributions at 4.8~GHz for the BL~Lacs and PR~QSOs are
drawn from the same population. 

Comparison of the histograms based on the data at 4.8~GHz with those at 14.5 
GHz shows a dramatic class-dependent difference. All the BL~Lacs in our
program exhibited significant amplitude changes (V $\geq 0.1$) in total flux
at {\it both} 14.5 and 4.8~GHz; for the QSOs the average amplitude of
variability is lower at 4.8~GHz. A KS test of the similarity of the distributions of V at 14.5 and 4.8~GHz for the BL~Lacs  indicates a probability of 0.377 that these distributions are drawn from the same population, quantifying the similarity of the distributions of the amplitude of variability
that we find at the two frequencies, while for the PR~QSOs it is only 0.022. This difference in behavior is a consequence of the fact that for the majority of events in the BL~Lac members the spectra in the cm-wavelength range remain
flat even during outburst development. Such behavior results from the generally low cm-wavelength opacity in the emitting regions during outbursts indicated by
our multifrequency flux data and contrasts with the behavior in many QSOs
where the amplitude of the variations, measured by V, is reduced at 4.8~GHz
({\it e.g.} Aller, Aller, Latimer, \& Hodge 1985), and where 
frequency-dependent time-delays in the outburst development on timescales of
years ({\it e.g.}  Hufnagel \& Bregman 1992) are often seen; we attribute such frequency-dependent behavior to self-absorption in the QSO emitting regions
 ({\it e.g.} Aller, Aller, \& Hodge 1981; Lobanov 1998a). The fact that we 
find few time-delayed, steep-spectrum events in  BL~Lac class members in spite of the large number of events followed during the nearly two-decade-long observing period, argues for differences in the physical conditions in the 
jets of BL~Lacs and QSOs. This result also supports the cm-to-submillimeter
study of Gear et al. (1994) which concludes that there is a class-dependent
difference in the underlying jet on the basis of the flatter spectra for
BL~Lacs than for QSOs in the 150-375~GHz range during relatively quiescent
phases. The interpretation that this is intrinsic to the source is 
further supported by the spectral indices inferred from radiative transfer modeling of BL~Lac itself (Hughes, Aller, \& Aller 1989b), and from an analysis
of the centimeter-to-millimeter spectral behavior of several
variable sources (Valtaoja et al. 1988), which finds values well below the
canonical values of $-0.6$ to $-0.7$. We, note, however, that there is currently little quantitative information available on position-dependent changes in opacity within the radio-emitting region of AGNs from spectral mapping. 

Figure 7 shows the variability index V(14.5) versus average cm-$\lambda$
spectral index. The range of average spectral index is typical of
that found for AGNs: almost all objects exhibit flat ($-0.5\le\alpha\le0.5$)
time-averaged spectral indices. The two sources outside this narrow range are
1413+135, an unusual red object ({\it e.g.} Bregman et al. 1981; Perlman et al.
1994) which exhibits a steeply-inverted spectrum during outbursts, and 2335+131
which has exhibited a spectrum that is characteristic of a transparent
emitting region (S$_{4.8}>$S$_{14.5}$) during almost the entire period of this study. For the PR QSO group, 7 of the 28 objects lie outside this range ($25\%$
of the sample compared to only $5 \%$ for the BL~Lacs).

Although our study finds variability differences between BL~Lacs as a group 
and QSOs and confirms the well-known relation between variability and flat 
radio spectra, the radio variability does not show a dependence on
distance (see Figure 8). Burbidge \& Hewitt (1992) and Valtaoja et al. (1992) have suggested that there may be more than one population of BL~Lac object, possibly with a separation at z$\sim0.6$ (Burbidge \& Hewitt 1992),
but we find that the low redshift objects are not readily distinguishable from the objects with high redshift on the basis of the amplitude of the
radio variability. Indeed, several of the most highly variable objects, including BL~Lac itself, are included in the low redshift group, and the
spread in V for objects with $0\leq$ z $\leq 0.1$ covers nearly the complete range. However, the 4 objects with z$\ge 0.9$ all exhibit high-amplitude variability with V $\ge 0.5$. The maximum correlation timescales of these
high redshift objects are also relatively short: they range from 0.9 to 2.6
years, while the timescales for the 38 BL~Lac for which values could be
determined range from 0.40 to 10.0 with $\bar \tau$=2.90. 

To search for possible correlations between the parameters presented in Table
1, we have carried out a cluster analysis ({\it e.g.} Nair 1997) for the variables z, S(14.5), V(14.5), V(4.8) $\alpha$, and $\tau$. The variables
have been standardized to allow for intercomparing the distributions of these
numerically, very-different parameters using:

\begin{equation}
x_{i}' = \frac{{x_i}-x_{av}}{\sigma}
\end{equation}
where $\sigma$ is the standard deviation and $x_{av}$ is the mean of the
sample. Both the Ward's minimum variance and single-link methods (using
the {\it JMP} Statistics package) have been applied. This cluster
analysis indicates that  the variability indices at 4.8 and 14.5~GHz are most strongly related and also identifies a weaker relation between $S_{av}$(14.5) and $\alpha_{av}$. In contrast with the result of Nair (1997), based on long-term Rosemary Hill optical data for a sample of 23 BL~Lacs (of which 20
are contained in our group) our cluster analysis results quantify that the amplitude of variability in the radio band does not increase with redshift.

\section{Linear Polarization}

Long-term polarization measurements provide unique information on the
characteristics of the flow. The average fractional polarization
provides a measure of the degree of ordering of the magnetic field in the emitting region, while the magnetic field orientation (orthogonal to the PA
orientation in a transparent synchrotron emitting region) combined with structural information from VLBI measurements can provide information on the flow direction ({\it e.g.} Aller, Aller, \& Hughes
1996b). The temporal behavior of the multifrequency polarization measurements
during active phases can be used to look for the signature of the passage of shocks in the flow, namely an increase in the degree of polarization and a temporal swing in the polarization position angle, resulting from an associated axial compression. Evidence in support of propagating shocks in the flows
through comparison of data with radiative transfer models ({\it e.g.} Hughes, Aller, \& Aller 1989b, 1991) and the identification of changes in jet-component orientation, plausibly interpreted as indicative of emerging shocks (Gabuzda
et al. 1994b), have already been presented for a few sample members. Results
of the analysis of the statistical properties of the polarization for the
source sample are discussed below.

\subsection {Average Percentage Polarization}

In Table~3 we tabulate polarization properties determined from this study
and summarize VLBI/P structural information taken from the literature. Columns
2 and 3 list adopted rotation measures and references. Faraday rotation in general is external to sources in our program objects ({\it e.g.} Jones \&
O'Dell 1977). For 1807+698 the value which minimized the frequency spread in 
our position angle determinations was selected from a range (Wrobel 1987, 1993) and may be local to the compact core (Wrobel 1987). In columns 4 through 
7 we tabulate the `representative' fractional polarization and its standard
deviation at 14.5 and 4.8~GHz; these were determined by averaging the
polarized flux found from monthly averages of the Stokes parameters Q and
U corrected for bias due to random noise, following the procedure of Wardle \&
Kronberg (1974). This parameter and the associated error is the most typical value of the fractional polarization and its spread and gives a measure of the expected value for a random measurement of the source. Because of the low
signal-to-noise in very weak sources ({\it e.g.} 2254+074 and 2335+031),
there is a bias towards including high values of the polarization in the 
determination of this parameter for these sources. While the position angle
of the electric vector of the polarized emission often exhibited large changes over the full time-period of the study, in several sources we could identify
relatively stable periods over the duration of several years, sometimes including several individual outbursts. In others, while the position angle varied widely, the majority of the measurements showed a preferred value when
all the measurements were considered. Further, some sources exhibited a
band of preferred values extending over several tens of degrees but avoided
the full range. To identify these preferred orientations in an unbiased way,
we adopted the following procedure. For each source a histogram of the
distribution of PA at each frequency based on 30-day averages of the data
(corrected for Faraday rotation) and binned by $10^{\circ}$ increments was
constructed to identify peaks indicative of  commonly occurring electric field
orientations;  only monthly averages $\geq3\sigma$ were included in the
construction of the histograms. For sources where the histogram exhibited a
significant deviation (probability of $\geq 99.5$\% as shown by $\sum\chi^2$) from a random distribution, the PAs within $30^{\circ}$ of those at the
 midpoint 
of the peaked bin were averaged: this both better-defined the peak and removed
contributions from possible secondary maxima in the distribution. The most
common intrinsic PA orientation for each source, $\chi_0$, determined in this way at 14.5 and 4.8~GHz is tabulated in columns 8 and 10 of Table~3. As a
measure of the dominance of the preferred orientation, we give in columns 9 
and 11 the fraction of 30-day periods when the PA measurements were within $\pm20^{\circ}$ of the most common value (and met the $3\sigma$ criterion on 
the data for inclusion). In computing this fraction, all averages were included in the total count  rather than only those with high S/N to prevent a bias towards times when the degree of polarization was highest. The fraction
tabulated thus represents  a lower limit to the fraction of averaged periods exhibiting the preferred orientation. Examples of the temporal behavior of the
position angle, the range of values of the Stokes parameters Q and U, and the
distribution of the polarization position angle found as described above are presented in Figures 9a-c and Figures 10a-c for AP~Lib and for 1308+326
respectively to illustrate the procedure. The detailed behavior of these
sources is discussed more fully below. To compare the orientation of the
magnetic field 
(the direction orthogonal to the electric vector PA orientation) 
with information on the parsec-scale flow indicated by VLBI/P maps, columns 12 and 13 give VLBI structural position angles taken from the literature or
provided privately, and their reference; two sources are unresolved as 
indicated by a `u' in column 12. The majority of the maps used to identify
source structure were made at 5 GHz; for many of our sample sources, we note
that jet curvature is known to be important both from these maps and from
comparison of small-scale (VLBI) and large-scale (VLA) structure. Also, we
note that in comparing structural and magnetic field orientations there is 
a $180^{\circ}$ ambiguity in the determination of the polarization position
angle so that our range of values is restricted to $0^{\circ}$--$180^{\circ}$ while the VLBI range is $0^{\circ}$--$360^{\circ}$.

Only seven sources exhibited a preferred intrinsic position angle orientation
in $\geq80$\% of the measurements at 4.8~GHz. Such behavior is illustrated in
Figure 9a which shows the data for AP Lib. Published structural information
for this source, which is difficult to map because of its southern location,
is very sparse and recent, so that structural changes during the time period
of our study have not been followed. The position angle at 4.8~GHz has
remained nearly stable with little spread  at an orientation $\sim 60^{\circ}$
from the VLBI structural axis. Note that there are no large changes in the polarization associated with the large total flux density event which peaked in 1994. The fact that we could identify a long-term stable period in so few objects is in marked contrast to the behavior we commonly see in QSOs ({\it e.g.} Aller, Aller, \& Hughes 1992) where preferred position angle orientations commonly persist over time periods including several events. {\it No} BL~Lacs exhibited comparable long-term stability at 14.5~GHz. 

In Figure~11 we show the distribution of the difference between the
polarization PA orientation and the VLBI structural axis, based on our
data at 4.8~GHz. A $\chi^{2}$ test indicates that the distribution does
not deviate significantly from a random distribution (p= 0.151); for
the distribution based on the 14.5~GHz data, the probability of no
deviation from randomness is even higher: 0.668. The result contrasts with 
that for the PR QSOs at 4.8~GHz where we have found a clustering near
$90^\circ$ (Aller, Aller, \& Hughes 1992). The fact that we do not find
a clustering near $0^\circ$, the preferred orientation found for BL Lac
jet components from VLBP studies, but rather that the distribution mimics that
found for the core components (Gabuzda \& Cawthorne 1992) suggests that the
integrated polarizations are dominated by emission from the unresolved cores.

While maximum degrees of optical polarization $\ge 20$\% are relatively
common in members of the BL~Lac class ({\it e.g.} Angel \& Stockman
1980), we have only rarely observed peak values $\ge 10$\%  in the
UMRAO monitoring program even during large outbursts; typical maximum
values are $<10\%$, and the average fractional radioband polarizations in 
BL~Lacs are generally only a few percent (See columns 4 and 6 in Table~3). In
Figures 12 and 13 we show the distribution of the average percentage
polarization found from averaging daily measurements of the Stokes parameters
Q and U for the BL~Lac sample and for the PR~QSOs at 14.5~GHz based
on measurements during the common observing period 1984.0-1996.7. This 
averaging procedure measures the average value from the temporal excursions
in the Q-U plane.  We interpret the low percentage polarizations that we find,
compared to the canonical values expected for synchrotron emitters, as 
evidence in support 
of turbulent magnetic fields in the radio-emitting regions (Jones et al. 
1985). The fact that we do not see a strong wavelength dependence of the
average polarization implies that Faraday depolarization is not responsible
for the low average percentage polarizations; the low degrees of polarization
from intra-knot (VLBI-scale) regions measured in a few cases and the inferred
absence of significant thermal material in the jet (Celotti et al. 1998) are consistent with this interpretation. The polarization distributions for the
two samples include the same range of values; this similarity suggests that
these two classes of objects have similar degrees of ordering of the magnetic
fields in their emitting regions. However, while for both classes of objects, 
the average polarization is $\le 4$\% for the majority of sample objects, 
a tail of higher polarization is apparent in the distribution for the BL Lacs;
A KS test indicates a probability of 0.032 that the distributions are drawn 
from the same population.

\subsection {Interpretation of Polarization Variability}

Variations in polarized flux and in position angle are common in the BL~Lacs
studied here, and few sources exhibited preferred orientations for of order
decade-long time intervals; this behavior is consistent with a scenario in
which the integrated polarization is dominated by time-dependent changes in
the turbulent flow, identified with evolving source components with lifetimes
of a few years to a decade in VLBI maps, rather than by 
the underlying quiescent jet. In two class members, BL~Lac and OT~081, whose
variability has been modeled with radiative transfer calculations for a shocked flow during the adiabatic-loss phase of the flare ({\it e.g.} Hughes, Aller, \& Aller 1989a, 1989b, 1991) monotonic swings in position angle orientation
accompanied by large changes in percentage polarization have provided evidence in support of the passage of a weak shock in the relativistic flows of the
jets. Evidence in support  of a shock-in-jet model has also been
provided by the success of such models in reproducing the broadband flux variability behavior in several sources (Marscher \& Gear 1985; Marscher,
Gear, \& Travis 1992; Litchfield et al. 1998). However, the emission is only
rarely dominated by the contribution from a single source component, at which
times it can show a clear signature of such behavior in our polarization
data -- a position angle flip and an increase in the fractional polarization; 
in these few cases
suitable for modeling, individual events are resolved in both the total and
polarized flux light curves. The best example is BL~Lac itself, for which the data are shown in Figure~14. The structural development of components in the VLBI maps has been followed by well-sampled measurements by Mutel and
coworkers ({\it e.g.} Mutel et al. 1990; Denn \& Mutel 1998); and the extrapolated onsets of new VLBI components match in time with the development 
of new activity in the total flux and linear polarization light curves ({\it e.g.} Aller, Aller, \& Hughes 1996b). Several events in the UMRAO data for 
this object have now been modeled; the results of that analysis show that
propagating transverse shocks can account for the behavior of some events, 
but the development of oblique shocks is required in others (Aller, Hughes, \&
Aller 1998). 

Evidence in support of a shock in the flow can also be seen in the data
for 0735+178 (Figure~15). The polarization swing and an increase in fractional 
polarization associated with the major flux event which commenced in 1988 can plausibly be interpreted as the result of a propagating shock. VLBP 
observations of 0735+178 in 1987.4 also suggest that a transverse shock
formed at this time, and the combined single-dish and VLBP measurements make a
convincing argument for this interpretation (Gabuzda et al. 1994b). When
periods of stable PA orientation occur in BL~Lac objects, these orientations
are often {\it along} or near to the VLBI structural axis such as we see in these two objects, indicating that the dominant magnetic field direction is 
{\it perpendicular} to the flow and consistent with the interpretation that a propagating shock has developed at these epochs. Such a preferred orientation can exist for many years and over many outbursts as is illustrated in Figure~16 for the intraday variable 2007+777 studied in the Bonn program ({\it e.g.} Krichbaum, Quirrenbach, \& Witzel 1992) which shows large excursions but a preferred position angle orientation nearly along the VLBI structural axis. 
This contrasts with the behavior in QSOs where the magnetic field, at least at 4.8~GHz, is typically aligned {\it along} the flow direction ({\it e.g.} Aller, Aller, \& Hughes 1992). 
 
\subsubsection {Noteworthy Exceptions to the Rule: Misclassified/transition objects}

Not all BL~Lacs show a preferred alignment of the magnetic field oriented
perpendicular to the flow direction during relatively stable periods. The
case of AP Lib has already been illustrated. Other exceptions, exhibiting
behavior typical of QSOs, are 1308+326 (Figure~10a) and 3C~446 (Figure~17). The 
position angle for 1308+326 has exhibited long-term preferred orientations
which flip between orientations parallel and perpendicular to the VLBI
structural axis. We have observed similar behavior in the QSO 3C~279 which
sometimes exhibits the orientation characteristic  of a BL~Lac ({\it e.g.}
Aller, Aller, \& Hughes 1996b). In the mid 1980s, during large events in both
total and polarized flux, the magnetic field was oriented perpendicular to the
flow direction, while in the late 1980s it was parallel to the flow direction.
VLBP maps during the relatively quiescent period 1987-89 also showed a 
QSO-like polarization orientation (Gabuzda et al. 1993). The data for 3C~446,
a source known to have extensive extended structure (Simon, Johnston, \& 
Spencer 1985), is shown in Figure~17; the position angle has remained
relatively stable near a value of $10\arcdeg$ since it was first observed at UMRAO in the late 1960s, except for the excursion in 1984 associated with a
decline to near 0\% polarization. The total flux light curve in this object exhibits the spectral evolution and frequency-dependent time delays characteristic of variations in self-absorbed sources; the large-amplitude but relatively infrequent outbursts seen here are also typical of QSOs rather
than BL~Lacs; and the dominant magnetic field direction is oriented near $100\arcdeg$ and aligned along the VLBI structural axis, as is more typically found for QSOs. Both the polarization and flux behavior suggest that this
object is either misclassed or that it is a transition object. 

\section{Origin of Class-dependent Differences}

Several pieces of evidence in the radio regime based on single-dish data have suggested that intrinsic differences exist between BL~Lacs and QSOs. They are: 1) the characteristic differences in the topology of the long-term position
angle data  described above; 2) results from an analysis of the evolution of
the Stokes parameters which delineated class differences based on the domain occupied in the Q-U plane (Hughes et al. 1994); and 3) modeling of specific
events in BL~Lac, OT~081, and the QSO 3C~279 based on radiative transfer
calculations which required relatively large viewing angles, on the order of
20$^{\circ}$-40$^{\circ}$, for all three objects (Hughes, Aller, \& Aller 1989; Hughes, Aller, \& Aller 1991). Viewing angles deduced independently from VLBI measurements for BL~Lac (Mutel  et al. 1990) and 3C~279 (Carrara et
al. 1993) agree well with the values determined from the modeling. The VLBI studies of Gabuzda and coworkers continue to support intrinsic class
differences ({\it e.g.} Gabuzda \& Cawthorne 1992). The class-dependent
spectral behavior of the total flux and  of the polarization orientation
and variability described in this work also point to dissimilarities 
between BL~Lac and QSO jets which we believe can plausibly be attributed to
differences in the characteristics of the parsec-scale flows of these objects.

Observations of the large-scale polarization structure of jets (Bridle 1984) have shown that the dominant field direction in their inner regions is parallel
to the jet axis. The long-term stability or relatively slow
changes in position angles that we find for many of the QSOs in the
UMRAO program, even over large changes in both total and polarized flux ({\it e.g.} Aller, Aller, \& Hughes 1996a), is consistent with the existence of a partially-ordered quiescent flow with shear along the jet axis, as proposed by Laing (1993, 1996) to explain the largescale (VLA) polarization structure. In BL~Lacs, where large-amplitude flux changes are the rule and rapid changes in polarization position angle are common, the contributions from evolving components, plausibly associated with shocks at least in
some cases, may dominate in determining the field direction. However, the
limited compression, viewing angle, and some opacity effects ensure that we
see at most only a modest increase in fractional polarization at any time -- so that our observed value is always strongly influenced by the emission from the 
underlying turbulent flow. Thus, while in both the case of the BL~Lacs and
of the QSOs we find low average fractional polarization, we attribute these to
different physical scenarios. In BL~Lacs, the occurrence of compressive
events is frequent, and the polarization is strongly influenced by evolving
components, although their effect on the fractional polarization is limited
as described above. In the case of the QSOs the polarization is dominated by
the contribution from the underlying quiescent jet where shear imposes a modest degree of order on the tangled magnetic field.

A plausible explanation for the differences that we find between QSO and
BL~Lac variability properties is that the bulk motions within BL~Lac
jets are slower than in QSO jets. The pattern speeds found from studies of VLBI components could provide a reasonable measure of the bulk motion in the flow since past work (Hughes, Aller, \& Aller 1989b, 1991) suggests that the 
shocks internal to the flow are relatively weak (Mach numbers for modeled outbursts
are in the range $1.11 \leq M \leq 1.52$) and perturb it very little. However, past statistical studies of VLBI component motions based on
different source groups, which attempted to identify class-dependent
differences in the distributions of $\beta_{app}$, have yielded conflicting results on whether these distributions for  BL~Lacs and QSOs are 
significantly different ({\it e.g.} Vermeulen 1996; Gabuzda et al. 1994a). Studies of samples containing larger numbers of BL~Lacs and based on temporally well-sampled data for the
unambiguous identification of evolving features, as well as higher-frequency
VLBI measurements ({\it e.g.} Lister, Marscher, \& Gear 1998), may be
required to resolve this important issue. Nevertheless, results from two dimensional hydrodynamical jet simulations including relativistic effects
already available ({\it e.g.} Duncan \& Hughes 1994; Komissarov \& Falle
1996; Mart\'i et al. 1995) are consistent with a scenario in which many of
the class-dependent characteristic properties described here can be accounted
for by differences in the flow speeds in the relativistic jets and the
resulting stability of the flow. The simulations support a picture in which
slower flows, plausibly associated with BL~Lac parsec-scale jets, are 
relatively unstable to the build-up of Kelvin-Helmholtz instabilities
(Hardee et al. 1998); hence, internal shocks can develop more readily in these
objects. Faster flows, plausibly associated with QSO jets, are relatively
stable, and there is little or no internal structure. The differences in
field topologies that we see in the radio polarization data and in the flux 
and polarization variability behavior could thus be explained as a consequence
of such speed-dependent stability. Theoretical studies permitting comparison 
of observations with  detailed simulations of the evolution of relativistic flows and of their synchrotron emissivity are clearly the best way to explore the dependence of  jet structure on bulk velocity and of the effects of introducing perturbations into the flow on its emissivity. Such studies are currently in very earlier stages of development ({\it e.g.} Mioduszewski,
Hughes, \& Duncan 1997; Gomez et al. 1997) and will provide a framework for
exploring the relation between observable VLBI structures and the underlying
flow as well as for understanding single-dish measurements.

The underlying cause of such proposed class-dependent differences in 
flow velocity is also not well-established. Unification schemes propose that
BL~Lac objects come from a parent population of FR~I class objects while QSOs are a subpopulation of FR~II sources ({\it e.g.} Urry \& Padovani 1995); and
while problems remain with accounting for some data within such a simple
scheme (e.g. Kollgaard et al. 1992, Wurtz et al. 1997), it provides a viable
explanation for many observed properties, including the class-dependent flow
differences discussed here. It has been suggested that differences in the accretion rates onto the putative black hole (Rees et al. 1982), generally
presumed to be the central engine of AGNs, and/or possibly differences in the
rotation speed of these black holes (Baum, Zirbel, \& O'Dea 1995), may result 
in collimated jet outflows which are different in FR~I and FR~II sources. 
This suggests an intrinsic difference in the central engine itself and its collimated outflow. 

\section{Discussion}

{\it All} objects in the UMRAO BL~Lac sample exhibited variability during the period of this study, and, as a group, the fractional variability
was higher in the BL~Lacs than in our QSO sample. Many of these BL Lacs
are among the most active objects in the UMRAO program, exhibiting nearly continuous flux changes with only short or no intermittent quiescent periods.

Almost all program objects have flat average spectra as expected for
variable sources. Observations which follow the temporal evolution of
individual events show that the variations in the majority of outbursts
are nearly simultaneous at all three of our frequencies and that the
spectra remain flat {\it even during outburst}. This suggests that the
spectral behavior is an intrinsic feature of the source and not due to opacity in a tapered flow. Our inhomogeneous synchrotron model fits for BL~Lac itself (Hughes, Aller,  \& Aller 1989b) support this interpretation; these required a
physical truncation of the flow, indicating that the inner edge of the jet was {\it not} the $\tau=1$ surface as is sometimes assumed when interpreting VLBI
maps. However, to fully explore the influence and range of opacity, spectral imaging providing maps of the turnover frequency in individual sources will be
required. Recent results for the QSO~3C~345 using this technique indicate that
the turnover frequency in the core is at 15 GHz, just at our highest frequency and is an order of magnitude lower in the extended regions of the source
(Lobanov 1998b); but no comparable detailed spectral maps of BL~Lac objects 
are yet published. If future spectral mapping information provides little or no
evidence for self-absorption in any part of the flow,  then we must conclude
that the visibility of the jet material might be determined, for example, by an initial phase of thermalization at a recollimation shock (Daly \& Marscher 1988); observational evidence in support of the `core' as a standing shock as expected with this scenario has recently been presented by Marscher (1998), for the well-studied QSO 3C~454.3, based on VLBA observations in the optically
thin regime.

In BL~Lacs, changes in polarized flux and position angle are common. While periods when a magnetic field orientation characteristic of a transverse
shock and associated compression can be identified in some events, the
distribution of long-term preferred PA orientation relative to the VLBI structural axis (flow direction) shows no preferred value. In QSOs the magnetic fields are often along the flow direction, and preferred long-term orientations are relatively common suggesting that the magnetic field is perturbed very
little during activity in these objects: weak shear is always more important 
in determining the field direction than the compression associated with
propagating shocks.

While our single-dish data have allowed us to identify the class-dependent spectral and variability properties of BL~Lac objects and to relate them to 
the flow conditions in the emitting regions, a combination
of such data with well-sampled VLBP measurements are required to 
fully study the effects of small-scale changes in the physical
conditions on the emission properties and to assess the importance
of such changes on the interpretation of the observed variability.
Past cm-wavelength VLBP maps at multiple epochs suggested that the dominant magnetic field direction is transverse in BL~Lac objects, but recent and
better-sampled VLBI data show that new components can emerge with a 
longitudinal field direction, possibly due to topological differences
in the local field direction in some cases (e.g. Gabuzda 1995) or,
alternatively, due to a range of shock orientations within the flow,
blurring this sharp division. Further, mm-wavelength VLBP maps (Lister,
Marscher, \& Gear 1998) are different from the centimeter-wavelength results, 
and are considerably more complex; this suggests that changes in the degree
of order with position within the radio-wavelength emitting region
probed by the waveband of observation may occur. To understand these effects,
field geometries during activity must be identified and followed to
investigate the relative importance of shocks themselves and of Doppler-related
changes, possibly associated with small-scale bends in the jets, in producing the variability. Coeval {\it multifrequency} VLBI/P data during such active periods should be used to identify topological differences, to trace spectral changes along the jet, and to identify position-dependent changes in the degree of ordering of the magnetic field within the radio-emitting region in the jet for an improved understanding of the conditions within the source-emitting regions.

\acknowledgments
We thank D. C. Gabuzda for helpful discussions, W. Xu  and D. C. Gabuzda for
making available unpublished data, M. C. Aller for assistance in data analysis
and preparation of the figures, and the NSF for partial support from grants
AST-9421979 and AST-9617032. This research has made use of the NASA/IPAC
Extragalactic Database (NED) which is operated by the Jet Propulsion Laboratory,
California Institute of Technology under contract with the National
Aeronautics and Space Administration, and of maps from the NRAO 15 GHz VLBA 
survey available via the NRAO website.

\clearpage



\clearpage

\begin{figure}
\caption{Two-week averages of the total flux density observations for
0235+164. Observations at 14.5, 8.0 and 4.8~GHz are denoted by crosses, 
circles, and triangles respectively. The complete time range of 
UMRAO data is shown for this source to illustrate the amplitude differences
between events in a single source. The large outburst in the mid 1970s
is not included in the analysis because it is outside the time period
studied here.
 \label{fig1}}
\end{figure}

\begin{figure}
\caption{Two-week averages of the total flux density observations for
ON 231. The symbols are as described in Figure~1.\label{fig2}}
\end{figure}

\begin{figure}
\caption{Histogram of V(14.5~GHz) for the BL~Lac sample. Results for those objects which sometimes exhibit QSO-like spectra are indicated by
cross-hatching. \label{fig3}}
\end{figure}

\begin{figure}
\caption{Histogram of V(14.5~GHz) for the QSOs in the Pearson-Readhead sample.
The objects have been separated into groupings on the basis of maximum-observed
polarization in the optical band as described in the text.  \label{fig4}}
\end{figure}

\begin{figure}
\caption{Histogram of V(4.8~GHz) for the BL~Lac sample. The hatching is
as in Figure 3. \label{fig5}}
\end{figure}

\begin{figure}
\caption{Histogram of V(4.8~GHz) for the  Pearson-Readhead sample with
separation by maximum-observed optical polarization.  \label{fig6}}
\end{figure}

\begin{figure}
\caption{V(14.5~GHz) versus cm-wavelength spectral index for the BL~Lac
sample. Triangles denote objects sometimes exhibiting QSO-like optical spectra.
\label{fig7}}
\end{figure}

\begin{figure}
\caption{V(14.5~GHz) versus redshift for the BL~Lac sample. Symbols are
as described for Figure 7. \label{fig8}}
\end{figure}

\begin{figure}
\caption{Figure 9a shows from bottom to top daily averages of the total flux density, the fractional polarization in percent, and the position angle of the
electric vector of the polarized emission for the very low redshift object
AP~Lib. The symbols are as in Figure~1. In this and subsequent plots of the
polarized emission, the polarization has been corrected for Faraday rotation using the rotation measures listed in Table~3. The dotted line in the top panel marks the {\it most common} value of the PA orientation. The range of fractional polarization exhibited by this object is typical of that for other class members. Figure 9b shows monthly averages of Q and U in percent at 4.8~GHz; the circle marks the typical fractional polarization determined from the averages
as described in the text; the arrow marks the orientation corresponding to the
most common polarization PA given in Table~3 for this source.  Figure 9c
shows the histogram of the PA distribution at 4.8~GHz. This is an example of a source exhibiting a nearly stable position angle during the duration of our
study. \label{fig9}}
\end{figure}

\begin{figure}
\caption{Figure~10a shows two-week averages of the total flux density,
the  polarized flux, and the polarization position angle for 1308+326. 
The symbols are as in Figure~1; the polarization has been corrected for
Faraday rotation. The horizontal line in the top panel marks the orientation
of the VLBI structural axis. Figures 10b and 10c show the Q-U and PA
distributions as described for Figure 9. This is an example of a source 
which shows two preferred PA orientations. \label{fig10}}
\end{figure}

\begin{figure}
\caption{Histogram of the difference between the VLBI structural
axis and the most common PA observed at 4.8~GHz. Objects exhibiting
a preferred orientation $< 50\%$ of the time are distinguished by
cross-hatching. \label{fig11}}
\end{figure}

\begin{figure}
\caption{Histogram of the average percentage polarization at 14.5~GHz 
for the BL~Lac sample. Objects which sometimes exhibit
QSO-like spectra are indicated by cross hatching. \label{fig12}}
\end{figure}

\begin{figure}
\caption{Histogram of the average percentage polarization at 14.5~GHz 
for the Pearson-Readhead sample with separation based on maximum-observed
polarization in the optical band. \label{fig13}}
\end{figure}

\begin{figure}
\caption{Two-week averages of the total flux density, the polarized flux and
the polarization position angle for BL~Lac. The symbols are as in Figure~1;
the polarization has been corrected for Faraday rotation. The line in the
upper panel marks the orientation of the VLBI structural axis.
\label{fig14}}
\end{figure}

\begin{figure}
\caption{Two-week averages of the total flux density, polarized flux and
polarization position angle for 0735+178. The symbols are as in Figure~1;
the polarization has been corrected for Faraday rotation. The line in the
top panel marks the orientation of the VLBI structural axis.\label{fig15}}
\end{figure}

\begin{figure}
\caption{Daily averages of the total flux density, the polarized flux and
the polarization position angle for 2007+777. The symbols are as in Figure~1;
the polarization has been corrected for Faraday rotation. The line in the
top panel marks the orientation of the VLBI structural axis. The source has exhibited very well-resolved events in total flux with rapid, well-defined
rises and declines; the emitting region is partial opaque during the
events in the late 1980s and early 1990s. Model-dependent estimates of the
viewing angle for this source from VLBI measurements are in the range
$16^{\circ}-21^{\circ}$ (Schalinski et al. 1992). \label{fig16}}
\end{figure}

\begin{figure}
\caption{Two-week averages of the flux and polarization for 3C~446. The symbols
are as in Figure~1; the polarization has been corrected for Faraday rotation.
Here the full time range of the UMRAO data is shown to illustrate
the stability of the PA over several decades of observation. The dashed line 
in the top panel marks the orientation {\it orthogonal} to the VLBI structural
axis indicated by the full line. In this source the magnetic field is aligned
nearly along the flow direction. \label{fig17}}
\end{figure}

\vfill\eject

\clearpage

\begin{deluxetable}{lrrrr}
\tablecaption{FLUX VARIABILITY INDICES. \label{tbl-2}}
\tablewidth{0pt}
\tablenum{2}
\tablehead{
\colhead{Group}      & \colhead{N} &
\colhead{Range} & \colhead{$V_{av}$} &
\colhead{$\sigma_{V}$} 
}
\startdata
BL~Lacs (14.5~GHz)  & 41 & $0.204 \leq V \leq 0.905$ & 0.555 & 0.174 \nl
PR~QSOs (14.5~GHz)  & 28 & $0.201 \leq V \leq 0.717$ & 0.381 & 0.145 \nl
BL~Lacs (4.8~GHz)   & 41 & $0.174 \leq V \leq 0.936$ & 0.501 & 0.163 \nl
PR~QSOs (4.8~GHz)   & 28 & $0.034 \leq V \leq 0.706$ & 0.266 & 0.155 \nl
\enddata
\end{deluxetable}

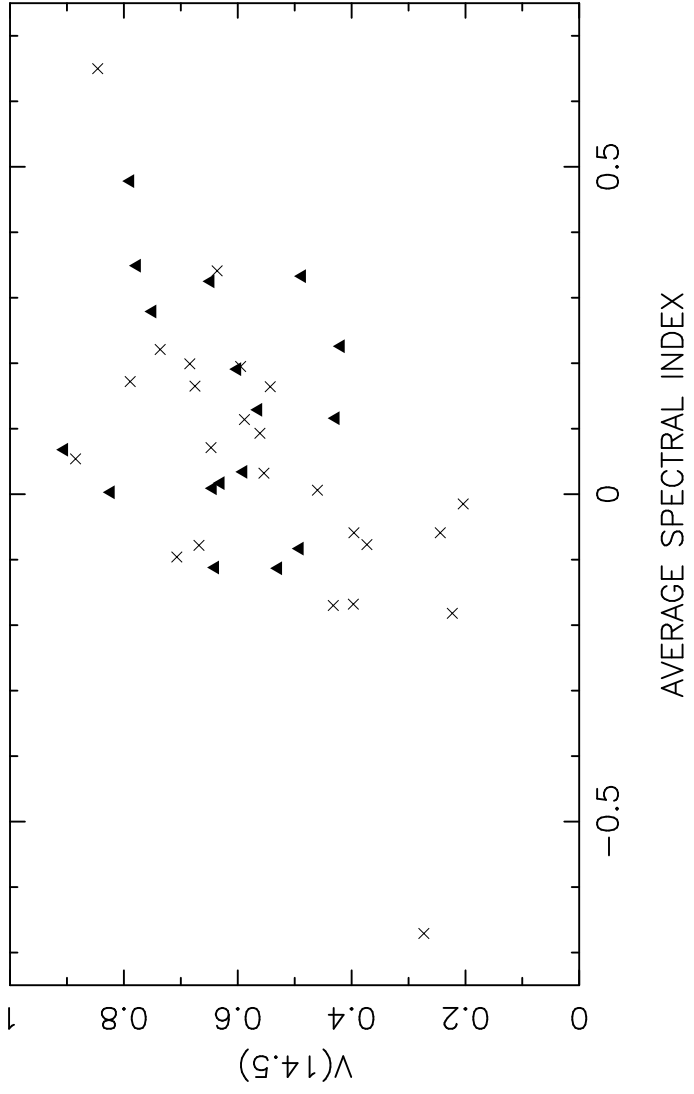
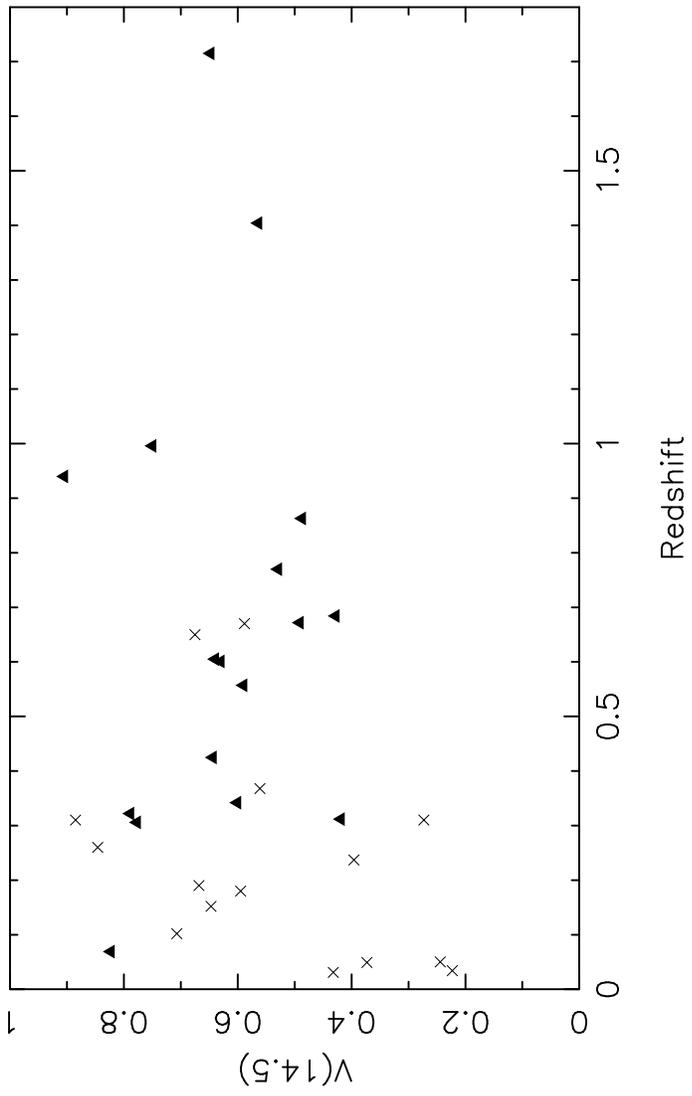
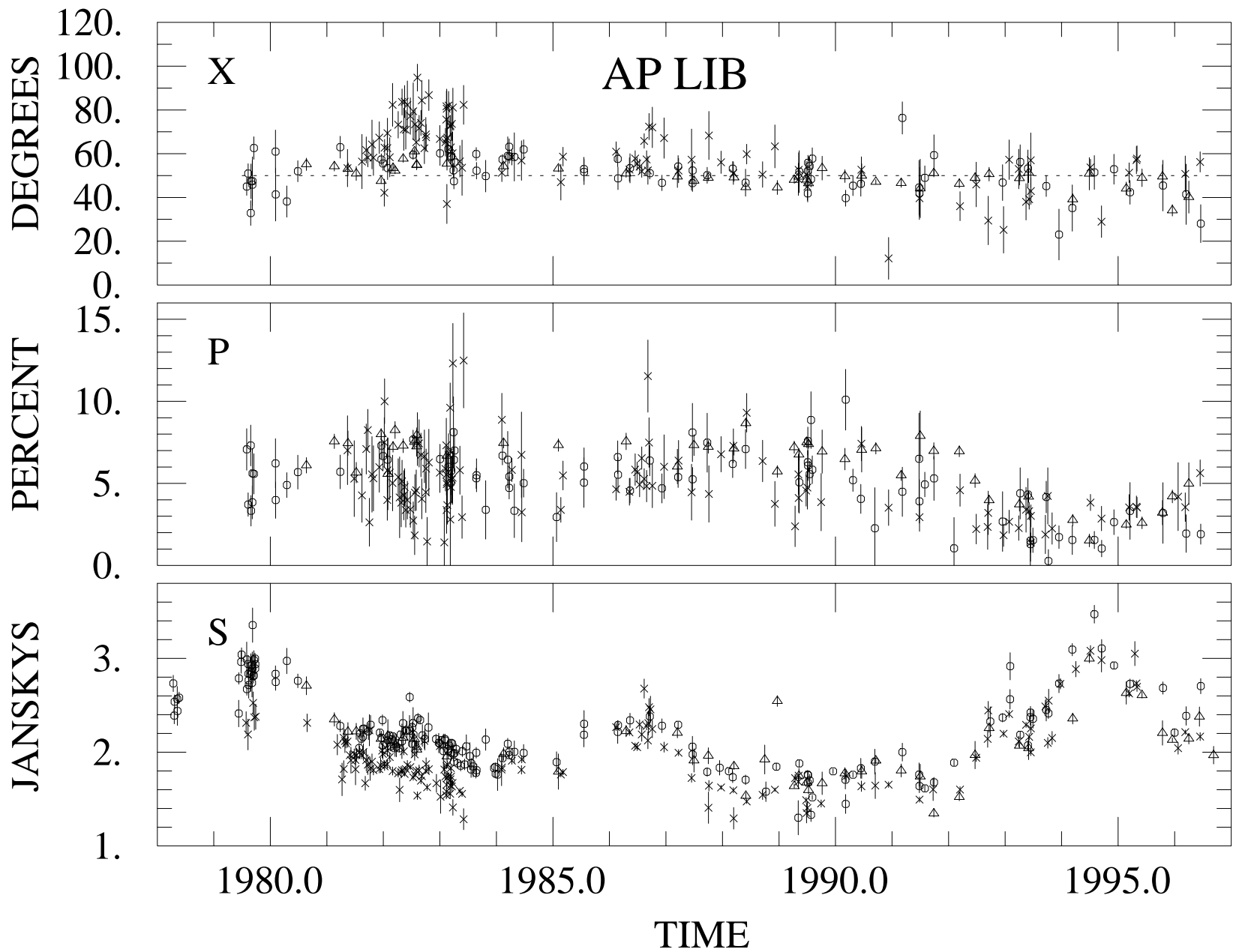
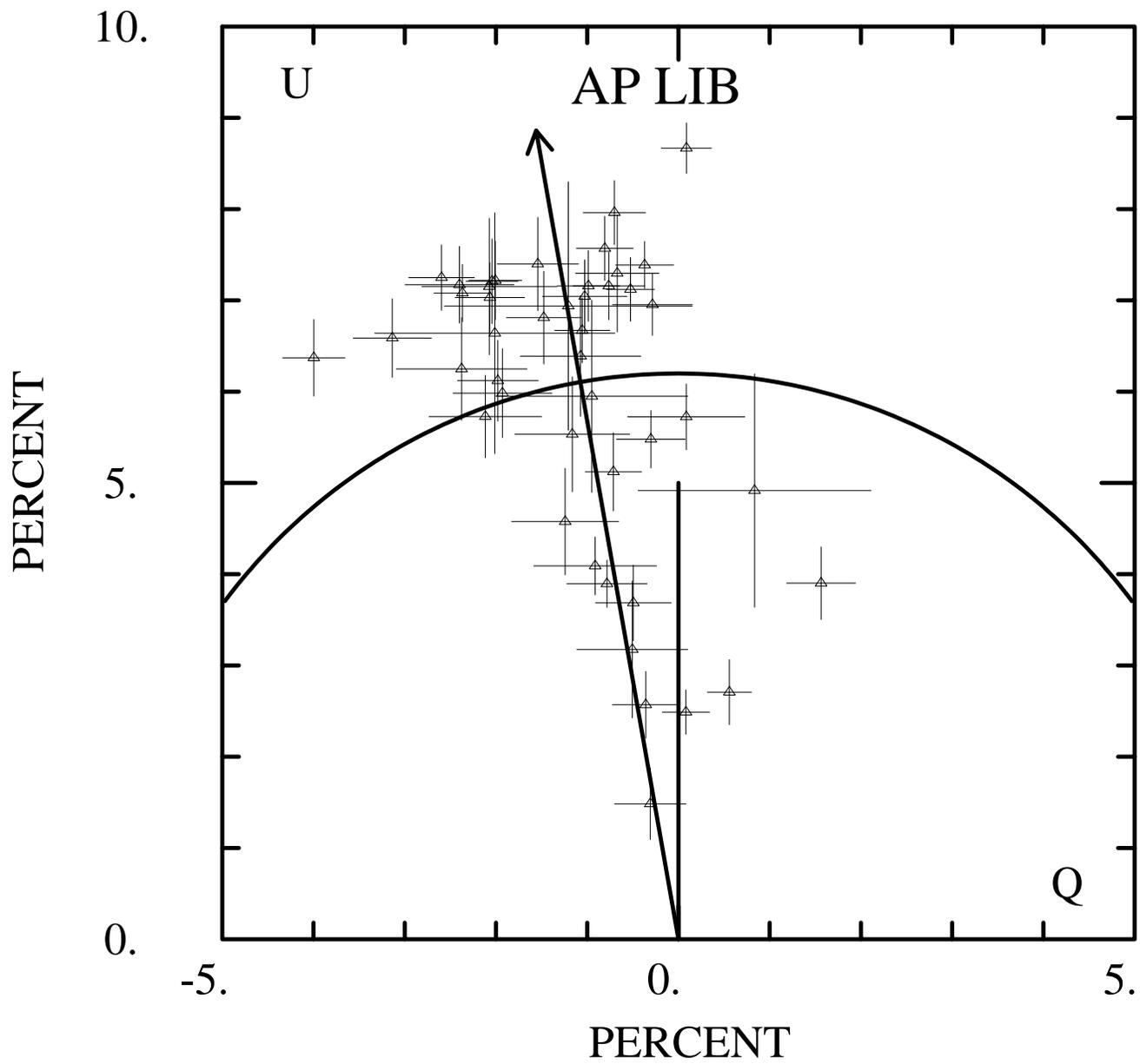
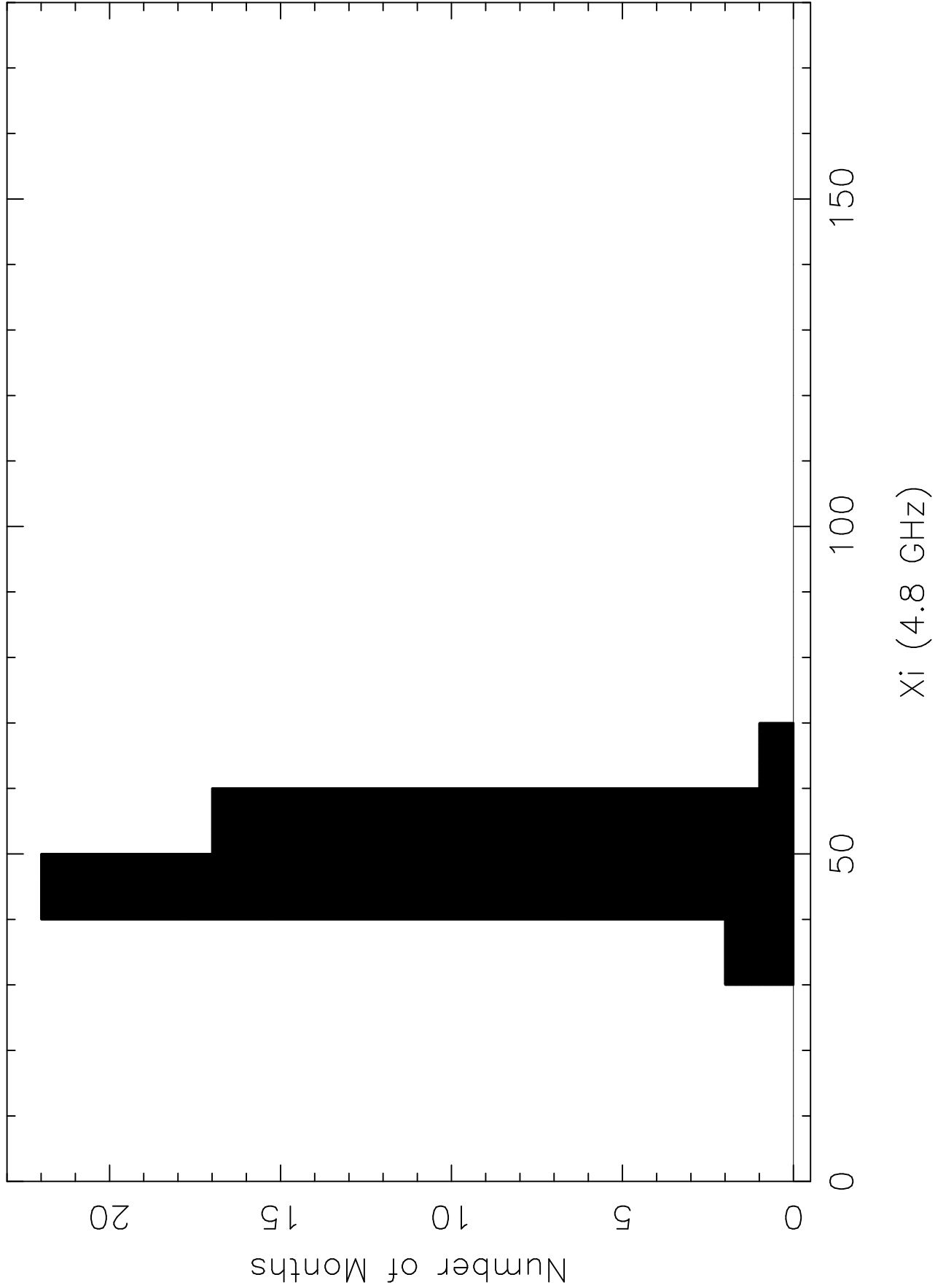
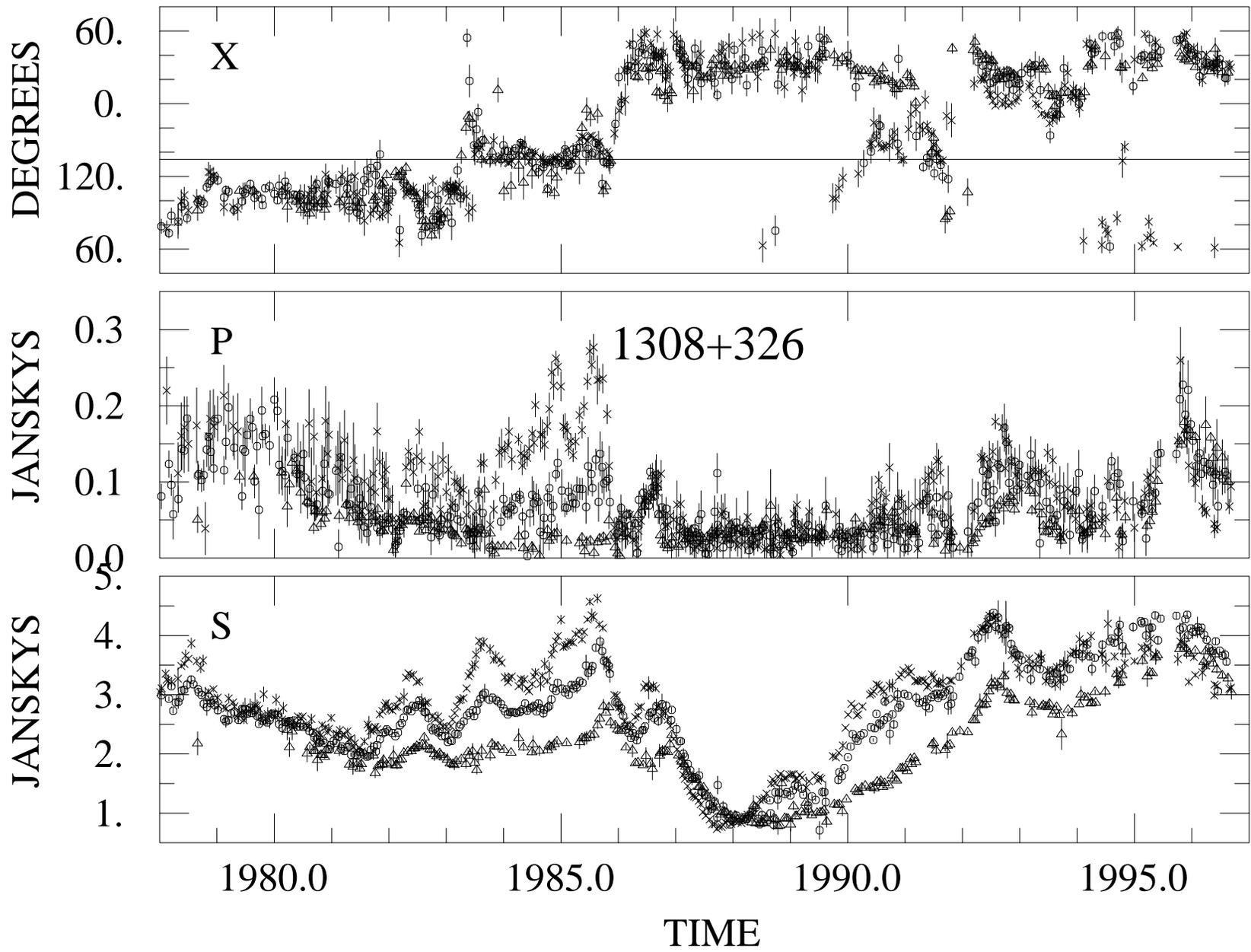
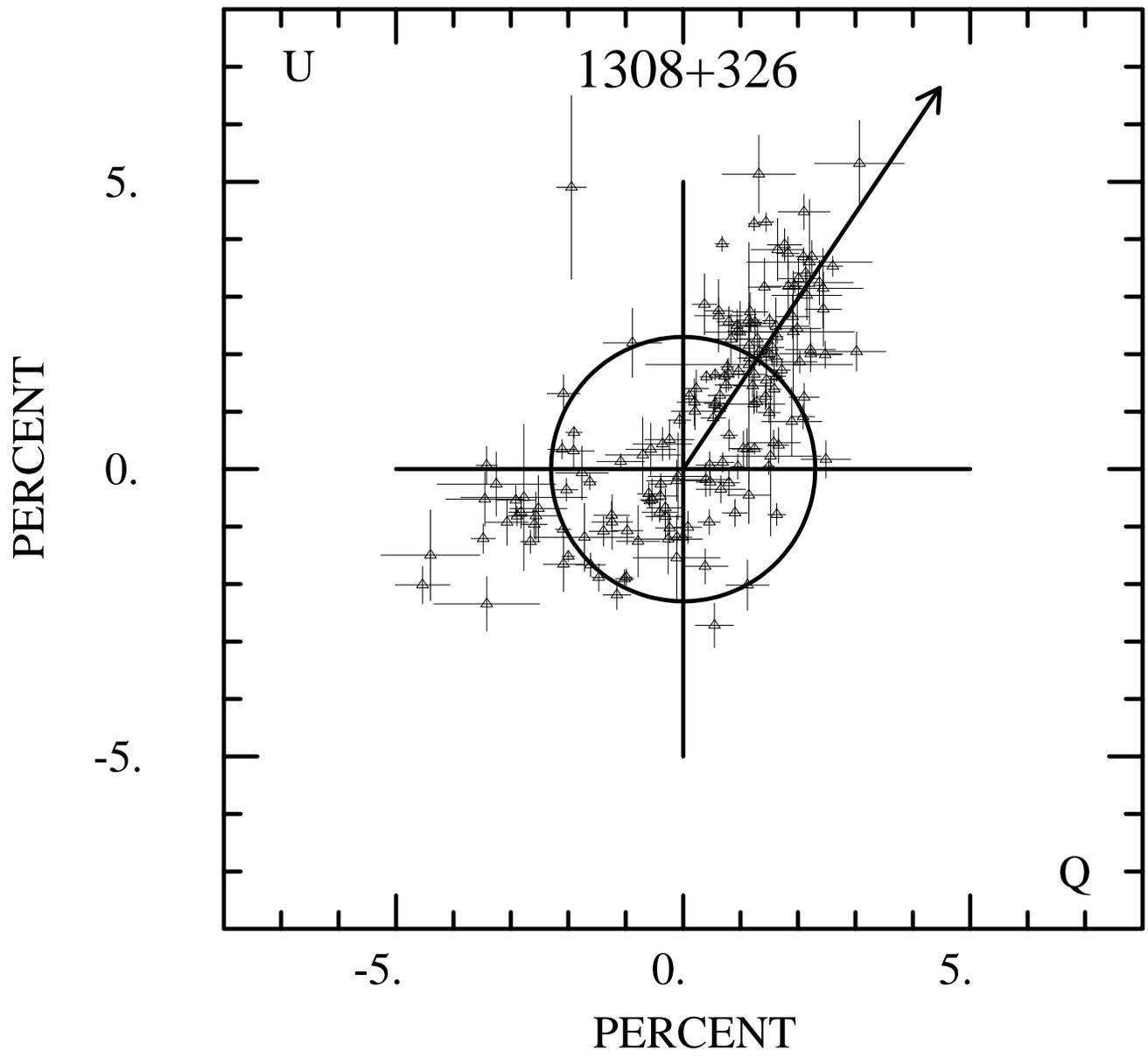
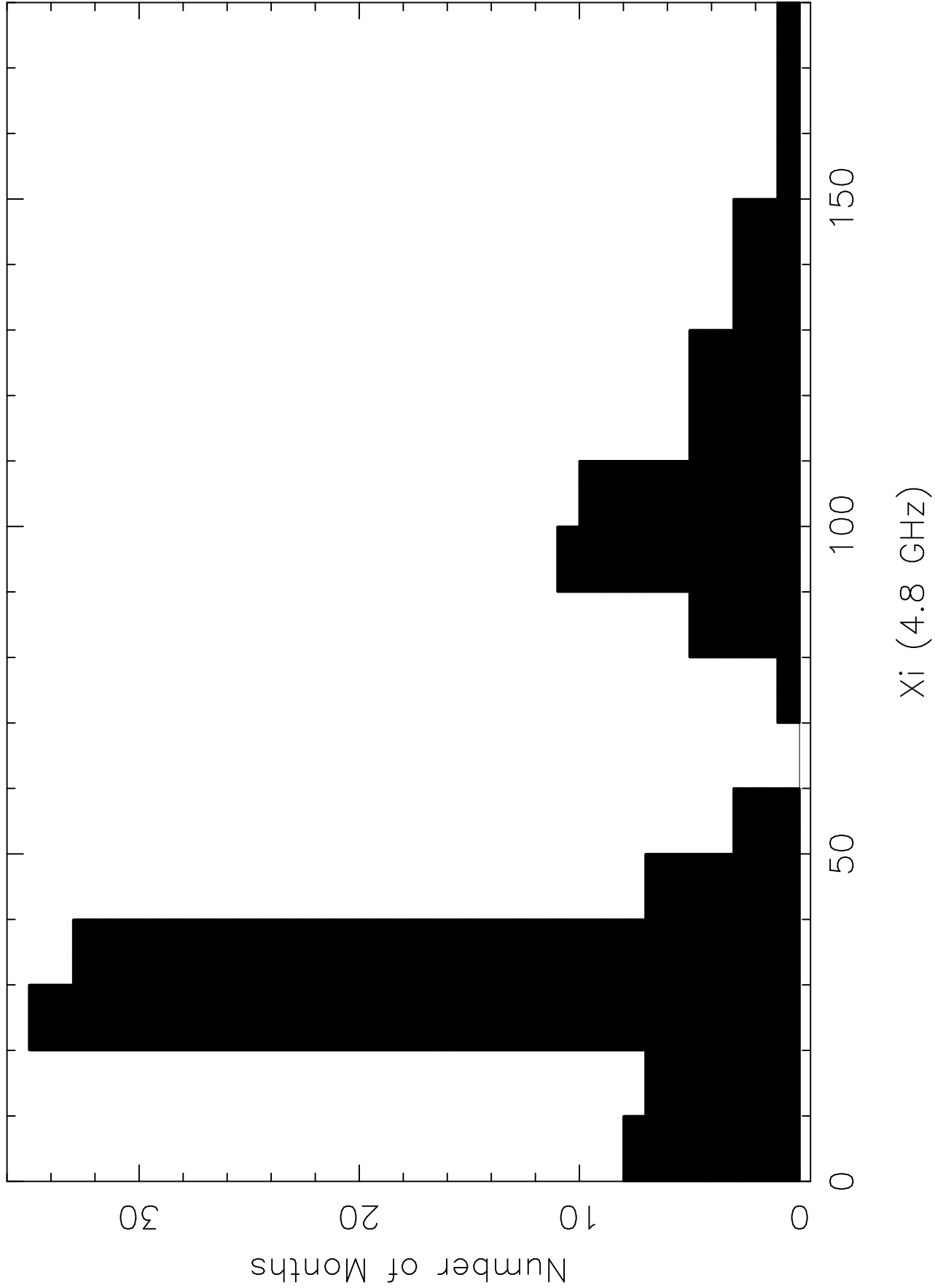
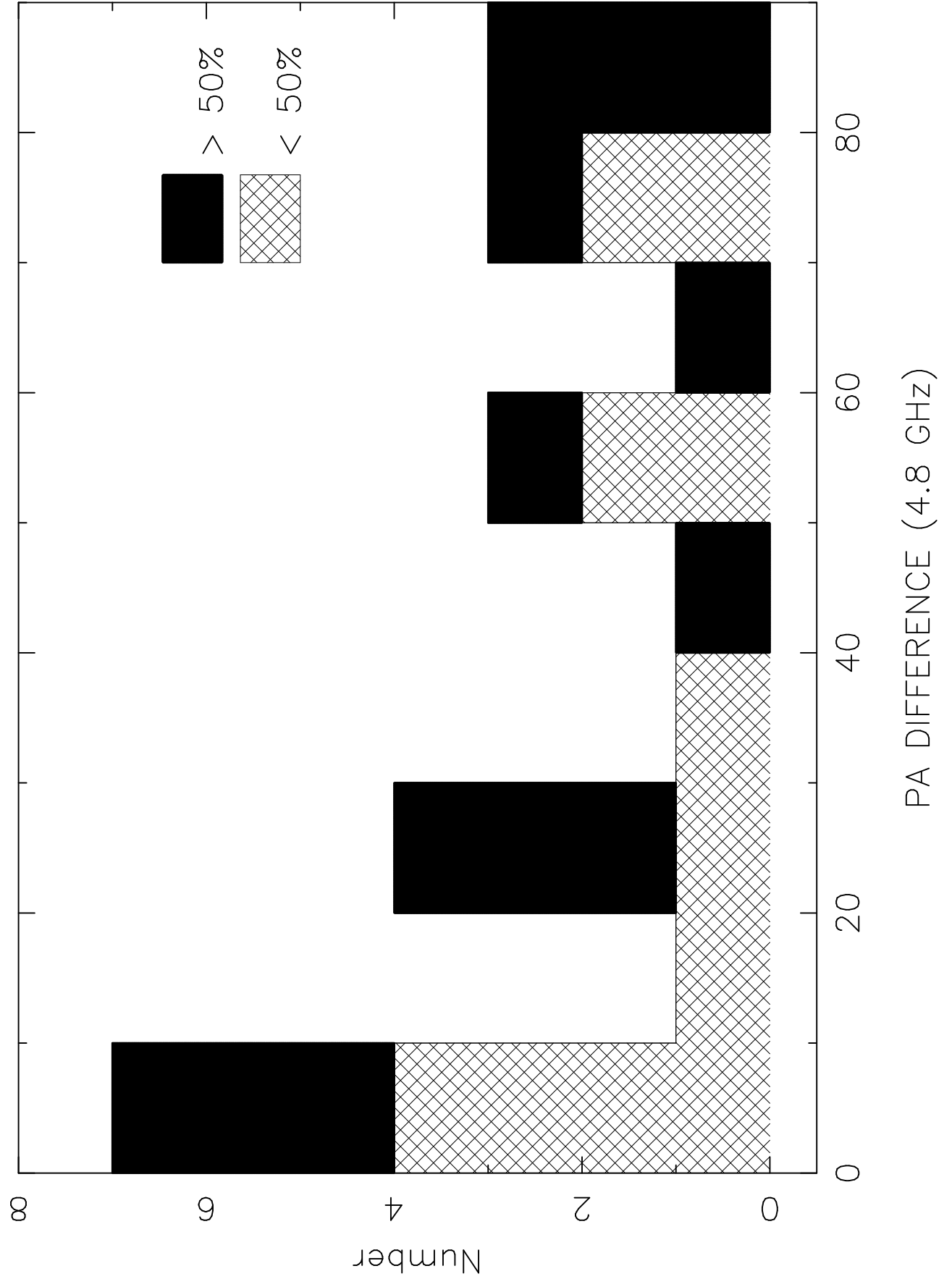
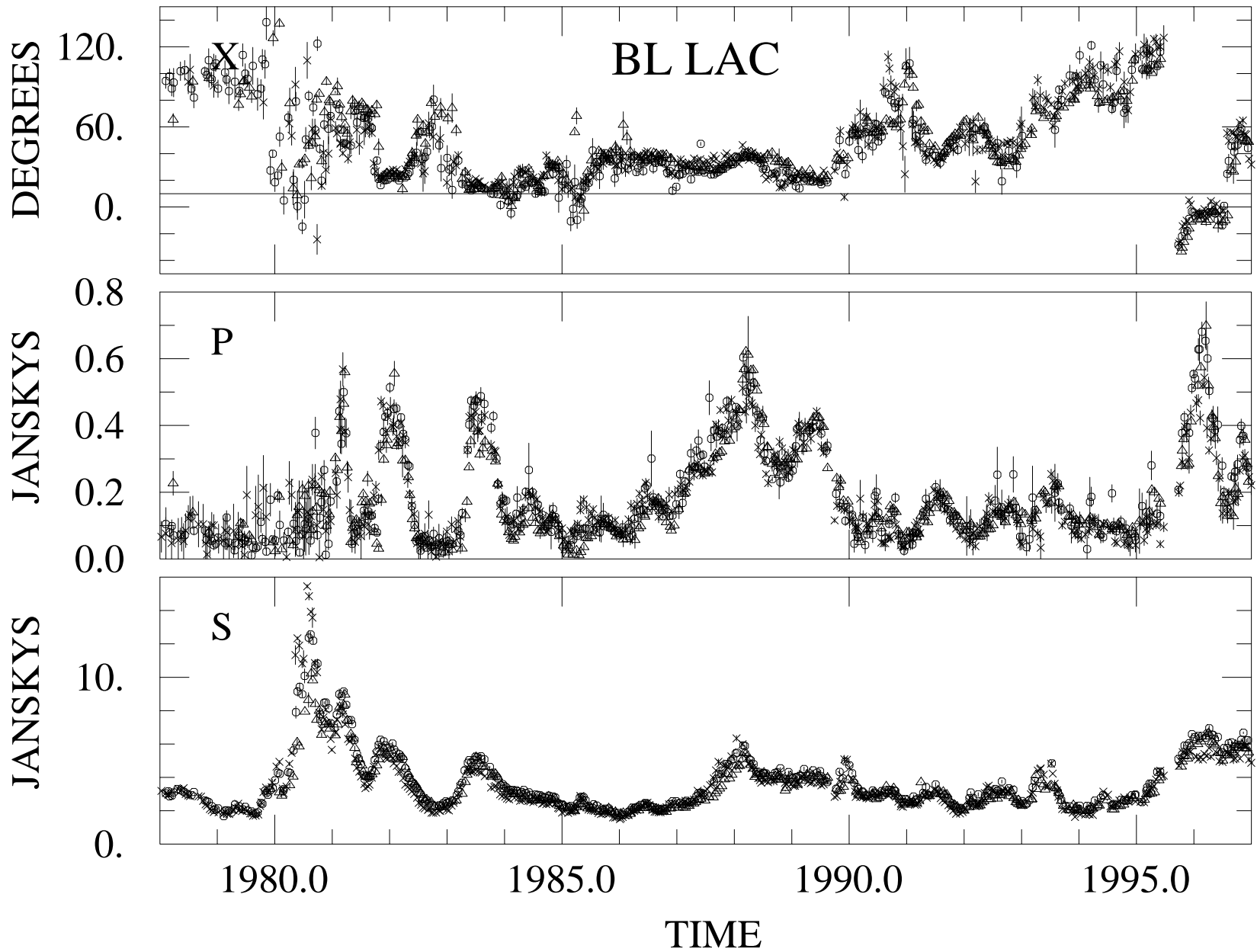
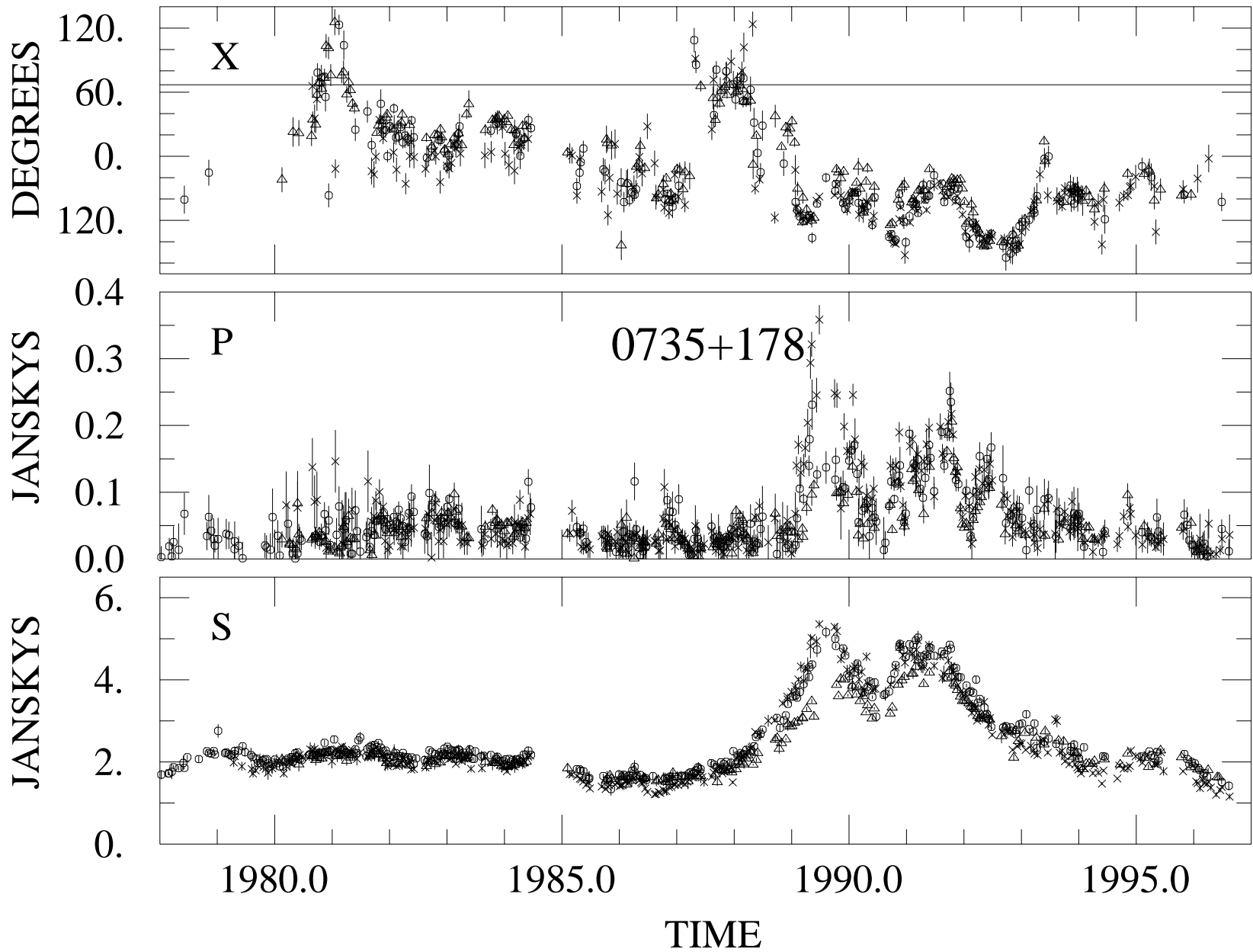
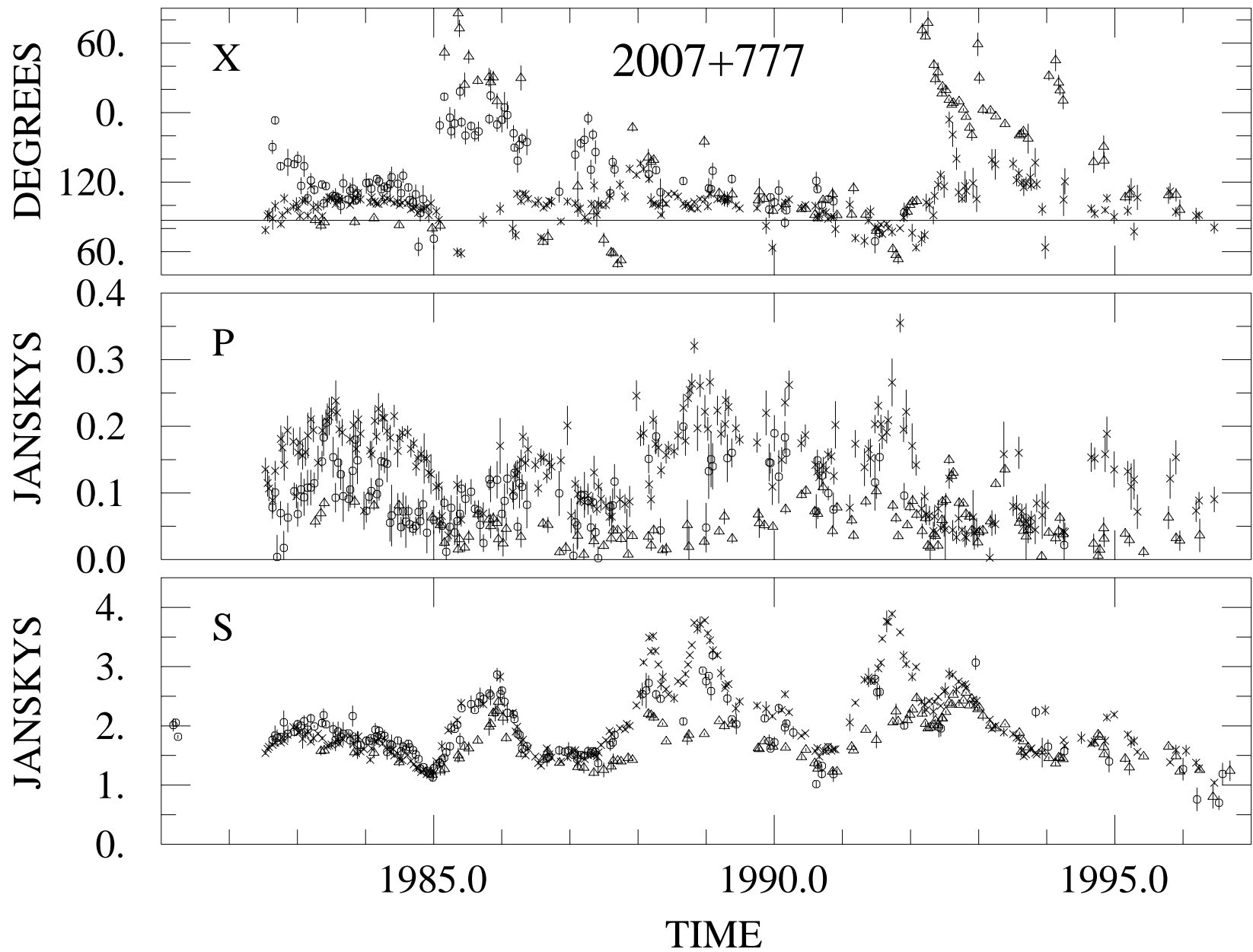
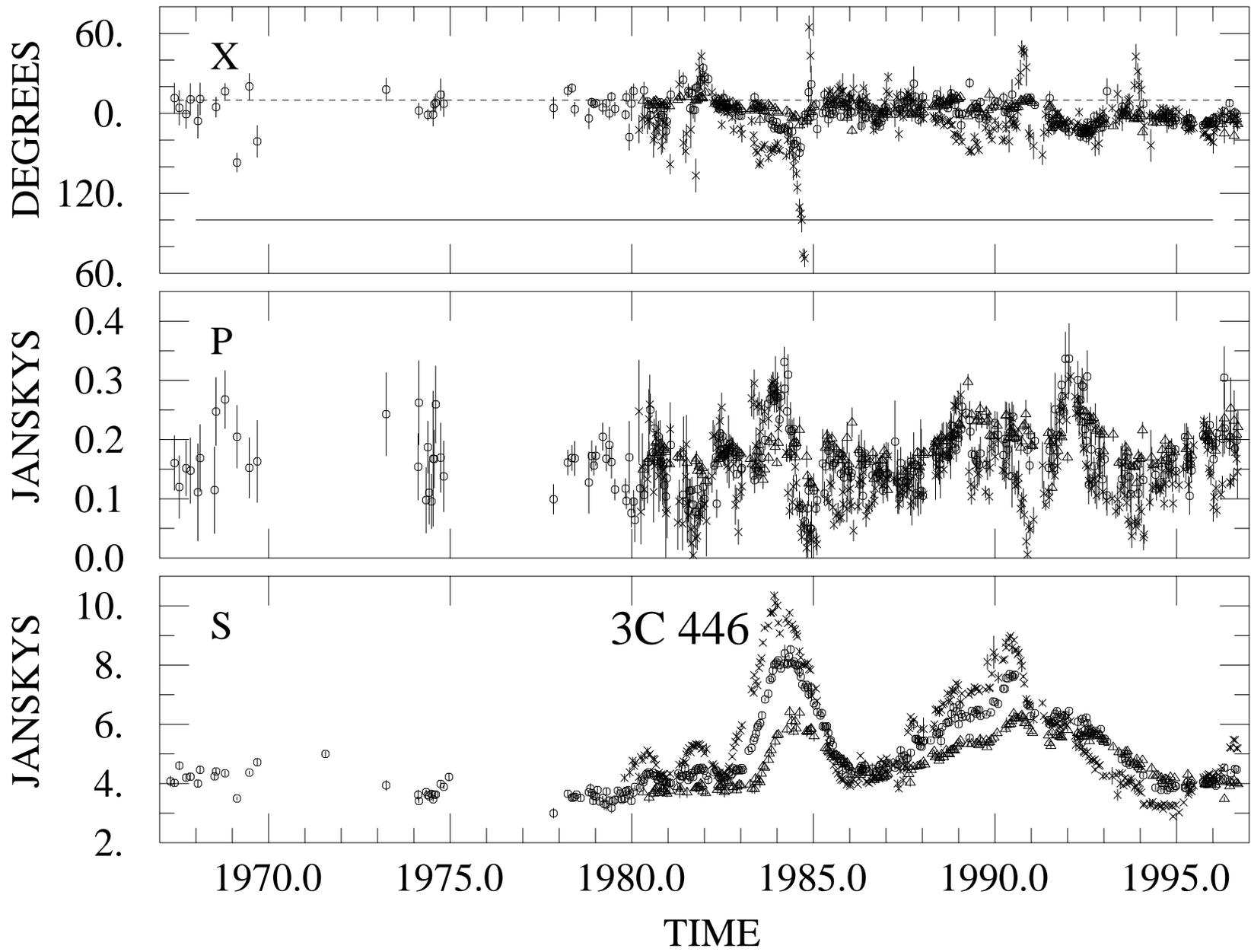
\end{document}